\documentclass[%
preprint,
 amsmath,amssymb,
]{revtex4-1}

\usepackage{graphicx}
\usepackage{dcolumn}
\usepackage{bm}

\begin{document}

\title{Manipulating orbital angular momentum entanglement by using the Heisenberg Uncertainty principle}
\author{Wei Li$^{1,3}$}
\email{alfred\underline{ }wl@njupt.edu.cn}
\author{Shengmei Zhao$^{1,2}$}
\email{zhaosm@njupt.edu.cn}
\altaffiliation{Nanjing University of Posts and Telecommunications, Institute of Signal Processing and Transmission, Nanjing, 210003, China}
\affiliation{$^{1}$Nanjing University of Posts and Telecommunications, Institute of Signal Processing and Transmission, Nanjing, 210003, China}
\affiliation{$^{2}$Nanjing University of Posts and Telecommunications, Key Lab Broadband Wireless Communication and Sensor Network, Ministy of Education, Nanjing, 210003, China}
\affiliation{$^{3}$Sunwave Communications CO., Hangzhou 310053, Zhejiang, People's Republic of China}

\date{\today}

 \begin{abstract}
Orbital angular momentum entanglement is one of the most intriguing topics in quantum physics. A broad range of research have been dedicated either to unravel its underlying physics or to expand the entanglement dimensions and degrees. In this paper, we present a theoretical study on the orbital angular momentum entanglement by employing the Heisenberg uncertainty principle to quantum position correlation within the azimuthal region. In this study, we decompose the pump light into a set of pump cone states characterized by their radii. The OAM entanglement can be manipulated by controlling the radius of the pump cone state, the length of the nonlinear crystal and the OAM carried by the pump field,which is followed by a detailed discussion. We expect that our research will bring us a deeper understanding of the OAM entanglement, and will do help to the high-dimensional quantum information tasks based on OAM entanglement.
\end{abstract}

\maketitle


\section{\label{sec:level1}Introduction}
\par A light beam with rotational symmetry carries a well defined orbital angular momentum(OAM), characterized by the winding number $l$ which ranges from $-\infty$ to $\infty$\cite{allen1992orbital}. In the process of spontaneous parametric down conversion(SPDC) in which a high-energy photon is converted into two photons with lower energy, the conservation law is not only fulfilled by energy and momentum, but also fulfilled by angular momentum(AM)\cite{arnaut2000orbital}. If there is no AM exchange between the nonlinear crystal and the incident photons, OAM conservation is fulfilled in SPDC, which gives rise to the generation of OAM entanglement\cite{walborn2004entanglement}. Since the first time it was discovered in experiment\cite{mair2001entanglement}, OAM entanglement has served as a promising candidate to accomplish a series of quantum tasks beyond two-dimensional Hilbert-space entanglement\cite{torres2011twisted}, for example, dense coding\cite{bennett1992communication}, high-dimensional teleportation protocol\cite{you2010schemes}, bit commitment\cite{langford2004measuring}, quantum cryptography\cite{groblacher2006experimental} and demonstration of high-dimensional entanglement\cite{fickler2012quantum,dada2011experimental,krenn2014generation}. Meanwhile, OAM entanglement has been successfully employed to explore some quantum features in experiments like violation of Bell's inequality for high-dimensional entanglement\cite{vaziri2002experimental,dada2011experimental,molina2007twisted}, quantum ghost imaging\cite{jack2009holographic} and EPR correlation between OAM and angular position\cite{leach2010quantum}. However, the experimentally generated OAM entanglement is far from being used directly in that the entanglement spectrum always has a finite bandwidth and the weight distribution for each mode is not uniform. Actual applications of OAM entanglement rely on the technique of entanglement concentration\cite{vaziri2002experimental,dada2011experimental,vaziri2003concentration}, in which both complex experimental procedures and good entanglement quality are required. Up to now, we still lack a thorough understanding  of OAM entanglement, and how to manipulate OAM entanglement remains an open question.

\par Almost all of the previous research works on OAM entanglement was based on mode coupling, i.e. the overlapping of the mode functions between the pump state and the signal and idler states\cite{franke2002two,torres2003quantum,osorio2008correlations,law2004analysis,PhysRevA.83.033816,Alison2011angular}. In these studies, the pump and the down-converted states are represented by LG modes, where the mode distribution is determined by the chosen radial modes\cite{torres2003quantum}. These kinds of entanglement should be called mode entanglement, which both cover azimuthal entanglement and radial entanglement. The total entanglement degree between the down converted photons is determined by the quantum correlation contributed from these two parts. Nonetheless, the angular position correlation in the azimuthal region has not been discussed yet, which is the real cause of the nonuniform distribution of OAM entanglement. Angular position and OAM are conjugate variables connected by Fourier transformation\cite{jha2008fourier}, and they form a EPR pair, the quantum correlation of one variable will determine that of the other\cite{Wei2018bell}. The first experimental study of mode distribution of OAM entanglement through angular position correlation is carried out by two-photon interference in the azimuthal domain\cite{pires2010measurement,jha2011partial,jha2011supersensitive}. It also has been shown in experiments that the increase in the dimension of OAM entanglement will lead to a stronger angular position correlation\cite{PhysRevA.86.012334}. Besides, angular position correlation is closely related to the radial coordinate, with all of the radial modes are contained in this two-photon interference experiment, the relationship between angular position correlation and OAM entanglement can not be truly reflected.

\par In this paper, we give a theoretical study of OAM entanglement in azimuthal domain and show that it is an inherent feature of angular-position entangled two-photon states generated by a rotational symmetrical pump beam. We decompose the pump light into a set of cone states characterized by its transverse momentum projection. The mode distribution of the OAM entanglement is determined by the radius of the pump cone state, the length of the nonlinear crystal and the OAM carried by the pump beam. The first two factors influence the OAM entanglement by controlling the quantum angular position correlation between the down-converted two-photon states, while the latter one modulates the OAM entanglement by shifting the diagonals of the OAM correlation spectrum.

\section{\label{sec:level1}Theory}

\begin{figure}
\centering
\includegraphics[width=10cm]{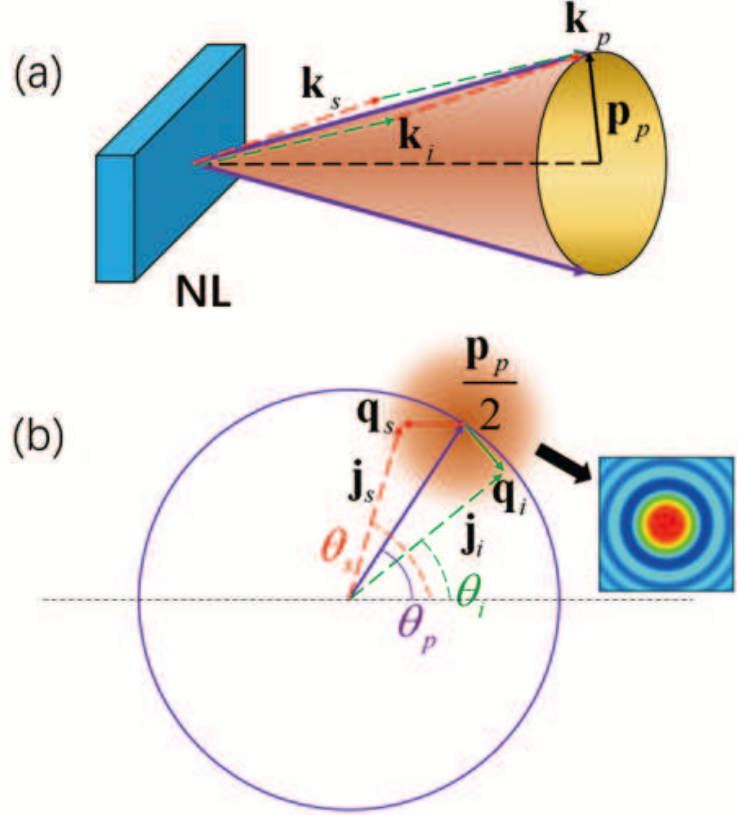}
\caption{Schematic illustration of the collinear SPDC. (a) The incident rotationally symmetric pump light beam is focused into the nonlinear crystal(NL). During the SPDC, the output of the pump state with a spherical phase on its wavefront can be decomposed into a set of cone states characterized by $\mathbf{p}_{p}$, the transverse momentum projection within the plane perpendicular to the propagation principal axis. (b) Transformation of the transverse momentum correlation from cartesian coordinate to polar coordinate. Inset: the schematic illustration of the down-converted two-photon transverse momentum correlation with respect to the pump momentum.}
\label{fig:Fig}
\end{figure}
\par We first consider an incident quasi-monochromatic plane-wave-like pump beam with a wave vector of $\mathbf{k}_{p}$ propagating through an uniaxial birefringent nonlinear crystal. The two-photon generation within the scheme of collinear type-$II$ energy-degenerate SPDC writes as\cite{hong1985theory}
\begin{equation}
    \left | \Psi  \right \rangle=\int \int d\mathbf{k}_{s}d\mathbf{k}_{i}
\Phi \left ( \mathbf{k}_{p},\mathbf{k}_{s},\mathbf{k}_{i} \right )
\left | \mathbf{k}_{s}\right \rangle \left | \mathbf{k}_{i}\right \rangle,
\end{equation}
where $\mathbf{k}_{s,i}$ are the wave vectors of the signal and idler states, respectively. The spectrum function $\Phi \left ( \mathbf{k}_{p},\mathbf{k}_{s},\mathbf{k}_{i} \right )$ determines the momentum correlation between the down-converted two photons. Here we omit the discussion of polarization during parametric conversion. $\Phi \left ( \mathbf{k}_{p},\mathbf{k}_{s},\mathbf{k}_{i} \right )$ arises from the phase-matching in SPDC, it has the following form\cite{hong1985theory}
\begin{equation}
\begin{split}
    \Phi \left ( \mathbf{k}_{p},\mathbf{k}_{s},\mathbf{k}_{i} \right )
=&\chi \left ( \omega _{p},\omega _{s},\omega _{i} \right )
\mathbf{E}_{p}\left ( \omega _{s}+\omega _{i} \right )\\
&\times \frac{\sin\left [ \frac{1}{2}\left ( \mathbf{k}_{p}-\mathbf{k}_{s}-\mathbf{k}_{i} \right )\cdot\mathbf{L} \right ]}{\frac{1}{2}\left | \mathbf{k}_{p}-\mathbf{k}_{s}-\mathbf{k}_{i} \right |}\\
&\times \exp \left [ i\frac{1}{2}\left ( \mathbf{k}_{p}-\mathbf{k}_{s}-\mathbf{k}_{i} \right )\cdot\mathbf{L} \right ],
\end{split}
\end{equation}
where $\mathbf{E}_{p}$ is the electrical field vector of the pump beam, $\chi\left ( \omega _{p},\omega _{s},\omega _{i} \right )$ is the bilinear susceptibility, $\omega_{p,s,i}=\left|\mathbf{k}_{p,s,i}\right|c/n_{p,s,i}$ are the central frequencies of pump, signal and idler photon states, $n_{p,s,i}$ are the corresponding refractive indices, $\mathbf{L}$ is the propagation vector of the pump beam within the nonlinear crystal, and the exponential term is the phase variances accumulated during the SPDC\cite{hong1985theory}.

\par It is a common feature that for the generation of OAM entanglement the pump state needs to be rotationally symmetric\cite{mair2001entanglement,torres2003quantum}. Schematic illustration of an OAM entanglement generation process is shown in Fig.1 (a), the plane-wave-like pump beam is focused onto a nonlinear crystal by a focal lens, and the output of the pump state with a spherical phase on its wavefront has a cone structure for its momentum distribution. Just as momentum comes from the translation symmetry in space, OAM comes from the rotational symmetry, thus it is better to study OAM entanglement in azimuthal region. In this scenario, we decompose the pump beam into a set of cone states labeled by $\left|\mathbf{p}_{p}\right|$, the transverse wave vector projection. Then the study of OAM entanglement is carried out by fixing the radial coordinate, and is thus only focused on the azimuthal region.

\par We first choose a pump state $\mathbf{k}_{p}$ on a pump cone $\left|\mathbf{p}_{p}\right|$. For a collinear SPDC, the conversion between the pump photon, the signal and idler photons can be divided into two directions, which are parallel and perpendicular to the wave vector of the pump, respectively. In this case, the parametric conversion mainly happens in the direction parallel to $\mathbf{k}_{p}$, while in the direction perpendicular to $\mathbf{k}_{p}$ there is no parametric conversion, otherwise the energy conservation law would be violated. Thus we have
\begin{equation}
    \left ( \mathbf{k}_{p}-\mathbf{k}_{s}-\mathbf{k}_{i} \right )\cdot\mathbf{L}=\left( k_{p}-k_{s}-k_{i} \right)L,
\end{equation}
where $L$ is the propagation length of the pump state $\mathbf{k}_{p}$ within the nonlinear crystal, $k_{p}$, $k_{s}$ and $k_{i}$ are the amplitudes of the longitudinal wave vector components along $\mathbf{k}_{p}$. In the paraxial approximation, these quantities write as follows\cite{molina2005control}
\begin{equation}
    \begin{split}
        k_{p}=& \left| \mathbf{k}_{p} \right |,\\
        k_{s(i)}=& \left | \mathbf{k}_{s(i)} \right| \left( 1-\frac{1}{2} \left| \frac{\mathbf{q}_{s(i)}}{\mathbf{k}_{s(i)}} \right |^{2} \right).
    \end{split}
\end{equation}
Here $\mathbf{q}_{s}$ and $\mathbf{q}_{i}$ are the transverse wave vectors for the signal and idler states.

\par By substituting equations (3,4) into (2), we obtain the two-photon correlation in the transverse wave vector representation as

\begin{equation}
      \begin{split}
          \Phi \left ( \mathbf{k}_{p}, \mathbf{q}_{s},\mathbf{q}_{i} \right )=&
          E_{p}L
\sin c \left [ \frac{1}{2}\left ( \frac{\left | \mathbf{q}_{s} \right |^{2}+\left | \mathbf{q}_{i} \right |^{2}}{\left | \mathbf{k}_{p} \right |} \right )L \right ]\\
&\times \exp \left [ i\frac{1}{2}\left ( \frac{\left | \mathbf{q}_{s} \right |^{2}+\left | \mathbf{q}_{i} \right |^{2}}{\left | \mathbf{k}_{p} \right |} \right )L \right ].
      \end{split}
\end{equation}
Equation (5) depicts the cone of the down-converted two photons centred at $\mathbf{k}_{p}$ with a radius $\Delta_\mathbf{q}$ defined as $\sqrt{\dfrac{2\pi \left|\mathbf{k}_{p}\right|}{L}}$, as shown in the inset of Fig.1 (b). For the pump state $\mathbf{k}_{p}$, the down conversion cone is totally determined by the thickness of the nonlinear crystal $L$.

\par The conversion from cartesian coordinates to polar coordinates is given in Fig.1 (b). Here $\mathbf{p}_{p}$, $\mathbf{j}_{s}$ and $\mathbf{j}_{i}$ are respectively the wave vector projections on the cross section of the pump cone for the pump, signal and idler states, while $\theta_{p,s,i}$ are the corresponding azimuthal angles. In the paraxial approximation, since $\mathbf{p}_{p}$ is far less than $\mathbf{k}_{p}$, the angle open by the pump cone state is so small that the cross section of the down-conversion cone is approximately parallel to the cross section of the pump cone. According to the trigonometric function calculation formula $\left|\mathbf{q}_{s,i}\right|^{2}=\left|\dfrac{\mathbf{p}_{p}}{2}\right|^{2}+\left|\mathbf{j}_{s,i}\right|^{2}-\left|\mathbf{p}_{p}\right|\left|\mathbf{j}_{s,i}\right|\cos\left(\theta_{s,i}-\theta_{p}\right)$, the spectrum function $\Phi \left ( \mathbf{k}_{p}, \mathbf{q}_{s},\mathbf{q}_{i} \right )$ can be transformed into:
\begin{widetext}
\begin{equation}
    \begin{split}
        \Phi \left ( \theta_{p},\theta _{s},\theta _{i} \right )=& E_{p}L
\sin c  \left [ \frac{\frac{1}{2}\left | \mathbf{p}_{p} \right |^{2}+\left | \mathbf{j}_{s} \right |^{2}+\left | \mathbf{j}_{i} \right |^{2}-\left | \boldsymbol{p}_{p} \right |\left | \mathbf{j}_{s} \right | \cos \left( \theta _{s}- \theta_{p}\right) -\left | \boldsymbol{p}_{p} \right |\left | \mathbf{j}_{i} \right | \cos \left(\theta _{i}-\theta_{p} \right) }{2\left | \mathbf{k}_{p} \right |}L \right ]\\
& \times \exp \left [i  \frac{\frac{1}{2}\left | \mathbf{p}_{p} \right |^{2}+\left | \mathbf{j}_{s} \right |^{2}+\left | \mathbf{j}_{i} \right |^{2}-\left | \boldsymbol{p}_{p} \right |\left | \mathbf{j}_{s} \right | \cos \left( \theta _{s}- \theta_{p}\right) -\left | \boldsymbol{p}_{p} \right |\left | \mathbf{j}_{i} \right | \cos \left(\theta _{i}-\theta_{p} \right) }{2\left | \mathbf{k}_{p} \right |}L\right ],
    \end{split}
\end{equation}
\end{widetext}
which represents the radial and azimuth correlations between the down-converted two photons in the polar coordinates. To simplify the discussion, we choose $\left|\mathbf{j}_{s}\right|=\left|\mathbf{j}_{i}\right|=\dfrac{1}{2}\left|\mathbf{p}_{p}\right|$. Now the down-converted two-photon state in the angular position representation reads
\begin{equation}
    \left | \Psi\left(\theta_{p}\right)  \right \rangle=\int \int d\theta _{s}d\theta _{i}\Phi \left (\theta_{p},\theta _{s},\theta _{i}\right ) \left | \theta_{s}  \right \rangle \left | \theta_{i}  \right \rangle.
\end{equation}
Here $\left| \Psi\left(\theta_{p}\right) \right\rangle$ represents the two-photon state converted from the pump state at angle $\theta_{p}$.

\par The Fourier relationship between angular position and OAM leads to\cite{jha2008fourier}
\begin{equation}
    \left| \theta \right \rangle=
    \frac{1}{\sqrt{2\pi}}\sum_{l=-\infty}^{\infty}
    \exp \left ( -il\theta \right)
    \left | l \right \rangle,
\end{equation}
with $\left|\theta \right\rangle$ the angular position state in the polar coordinates, and $\left|l\right\rangle$ the OAM state. In the angular position representation, the pump cone state carrying an OAM of $l_{p}$ reads
\begin{equation}
    \left | \Phi_{p} \right \rangle=
    \frac{1}{\sqrt{2\pi}}\int d\theta_{p}
    \exp \left( il_{p}\theta_{p} \right)
    \left | \theta_{p} \right \rangle.
\end{equation}
 Here $\left | \Phi_{p} \right \rangle$ is the pump cone state, $\left | \theta_{p} \right \rangle$ is the angular position state, the amplitude of the pump state distributes uniformly in the range from 0 to $2\pi$. By substituting equation (7), (8) into (9), the quantum correlation of the down-converted two-photon state pumped by a cone state in the OAM representation is
\begin{equation}
    \begin{split}
        \left | \Psi \right \rangle =&\frac{1}{\sqrt{2\pi}}\int d\theta_{p} \exp\left(il_{p}\theta_{p}\right)\left|\Psi\left(\theta_{p}\right)\right\rangle\\
        =&\frac{1}{2\pi\sqrt{2\pi}}
    \sum_{l=-\infty}^{\infty}\sum_{l'=-\infty}^{\infty}A\left( l_{s},l_{i}\right ) \left | l_{s} \right \rangle \left | l_{i} \right \rangle\\
    &\int d\theta_{p}\exp \left [ i\left( l_{p}-l_{s}-l_{i} \right) \theta_{p} \right]\\
    =&\frac{1}{2\pi}\sum_{l_{s}=-\infty}^{\infty}\sum_{l_{i}=-\infty}^{\infty}
    A\left( l_{s},l_{i}\right ) \delta_{l_{s},l_{p}-l_{i}}\left | l_{s} \right \rangle \left | l_{i} \right \rangle\\
    =&\frac{1}{2\pi}\sum_{l=-\infty}^{\infty}A\left( l,l_{p}-l\right )\left | l \right \rangle \left | l_{p}-l \right \rangle.
    \end{split}
\end{equation}
In equation (10), $A\left(l_{s},l_{i}\right)$ is a two dimensional quantum OAM correlation spectra for the signal and idler states with a form of
\begin{equation}
    \begin{split}
    A \left( l_{s},l_{i} \right)=&\int \int d\theta_{s} d\theta_{i}
    \phi \left( \theta_{p},\theta_{s},\theta_{i} \right)\\
    &\times \exp \left [ -il_{s} \left(\theta_{s}-\theta_{p}\right)-il_{i}\left(\theta_{i}-\theta_{p}\right)\right],
    \end{split}
\end{equation}
which is a 2$D$ Fourier transformation of $\phi\left( \theta_{p},\theta_{s},\theta_{i} \right)$. From equation (10), we can see that OAM entanglement in SPDC arises from the continuous rotational symmetry of the pump state, where the entanglement spectrum $A\left(l,l_{p}-l\right)$ is one of the diagonals of $A\left(l_{s},l_{i}\right)$ shifted by $l_{P}$ from the zero point. Thus pumped by a rotationally symmetric light beam, the two-photon OAM entanglement is totally determined by $A\left(l_{s},l_{i}\right)$, which is a conjugate part of $\phi\left( \theta_{p},\theta_{s},\theta_{i} \right)$. In the practical SPDC experiments conducted on the nonlinear crystal with a specified cut angle, and perfect phase-matching is guaranteed within a small angle range, tunable parameters are the radius of the pump cone state $\mathbf{p}_{p}$, the length of the nonlinear crystal $L$ and the OAM of the pump state $l_{p}$. In the following, we will give a detailed study of how these parameters influence the OAM entanglement.

\begin{figure}
    \centering
    \includegraphics[width=10cm]{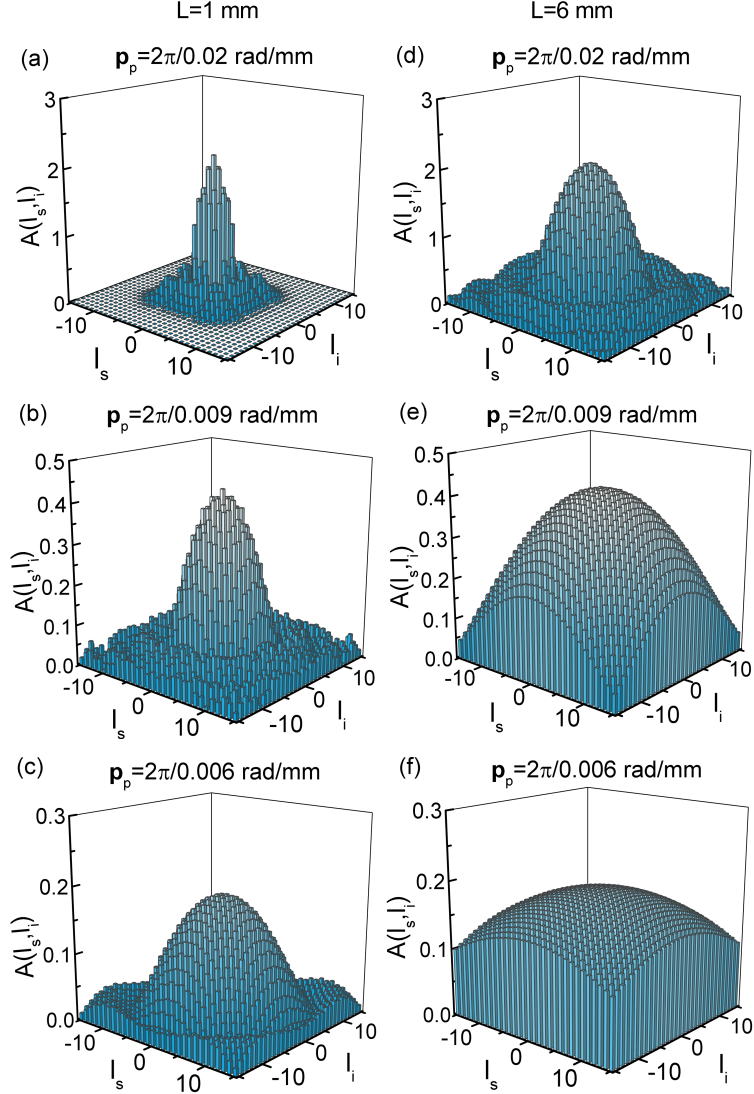}
    \caption{Evolution of the two-photon OAM quantum correlation spectrum $A\left( l_{s},l_{i} \right)$ with respect to the radius of the pump cone state, $\left| \mathbf{p}_{p} \right|$ at $L$= 1 $mm$ (a-c) and $L$ = 6 $mm$ (d-f).}
    \label{fig:my_label}
\end{figure}

\section{\label{sec:level2} Results and Discssion}
\par Figure 2 shows the simulation results of the evolution of the OAM correlation spectra $A\left(l_{s},l_{i}\right)$ with different values of $\left|\mathbf{p}_{p}\right|$ and $L$. The pump wavelength is 400 $nm$ and the wavelengths of the signal and idler states are both equal to 800 $nm$. In Fig.2 (a-c), the nonlinear crystal length $L$ is 1 $mm$, and in Fig.2 (d-f), $L$ is 6 $mm$. $A\left(l_{s},l_{i}\right)$ is a weak two-dimensional sinc-like function of $l_{s}$ and $l_{i}$ peaking at $l_{s,i}=0$. According to Fig.2, the bandwidth of $A\left(l_{s},l_{i}\right)$ increases with $\left|\mathbf{p}_{p}\right|$ and $L$. This phenomena can be explained by the Heisenberg uncertainty principle. In a collinear SPDC with a pump state of $\mathbf{k}_{p}$, both $\left|\mathbf{q}_{s}\right|$ and $\left|\mathbf{q}_{i}\right|$ are confined within a small down conversion cone (see equation (5)). Here we choose an appropriate pump cone state in which $\left|\mathbf{q}_{s(i)}\right|\ll \left|\mathbf{p}_{p}\right|/2$, then we'll get $\cos \left ( \theta _{s,\left ( i \right )}-\theta _{p} \right )=1-\frac{1}{2}\left ( \theta _{s,\left ( i \right )}-\theta _{p} \right )^{2}$. Furthermore, by choosing $\left|\mathbf{j}_{s(i)}\right|=\left|\mathbf{p}_{p}\right|/2$, equation (6) writes

\begin{equation}
    \begin{split}
        \Phi \left ( \theta _{p},\theta _{s},\theta _{i} \right )\approx  &E_{p}L
\sin c  \left[\frac{\left|\mathbf{p}_{p}\right|^{2}\left(\left(\theta_{s}-\theta_{p}\right)^{2}+\left(\theta_{i}-\theta_{p}\right)^{2}\right)}{8\left|\mathbf{k}_{p}\right|}L\right]\\
&\times\exp\left[i\frac{\left|\mathbf{p}_{p}\right|^{2}\left(\left(\theta_{s}-\theta_{p}\right)^{2}+\left(\theta_{i}-\theta_{p}\right)^{2}\right)}{8\left|\mathbf{k}_{p}\right|}L\right].
    \end{split}
\end{equation}
By choosing $\theta_{p}=0$, the angular distribution of the down-converted two-photon state is confined within a small angle range with a width of
\begin{equation}
    \Delta\left(\theta_{s}^{2}+\theta_{i}^{2}\right)=\frac{8\pi \left|\mathbf{k}_{p}\right|}{\left|\mathbf{p}_{p}\right|^{2}L}.
\end{equation}
Here $\Delta\left(\theta_{s}^{2}+\theta_{i}^{2}\right)$ can be viewed as the uncertainty in the relative angle distribution of the signal and idler states on the pump cone, and it is inversely proportional to $L$ and the square of $\left|\mathbf{p}_{p}\right|$. As a result, the larger the pump cone radius $\left|\mathbf{p}_{p}\right|$ is (the longer the crystal length $L$ is), the more certain we can know about relative angular position distribution of the down-converted two-photon states. According to Heisenberg uncertainty principle, for two conjugate variables, an increase in the certainty of one variable will lead to a decrease of the certainty in the other. Here it is found that the uncertainty principle is also applicable to two-part system. Therefore, by increasing the relative angular position correlation of the down-converted two-photon state, we will get a larger bandwidth of $A\left(l_{s},l_{i}\right)$.

\begin{figure}
    \centering
    \includegraphics[width=15cm]{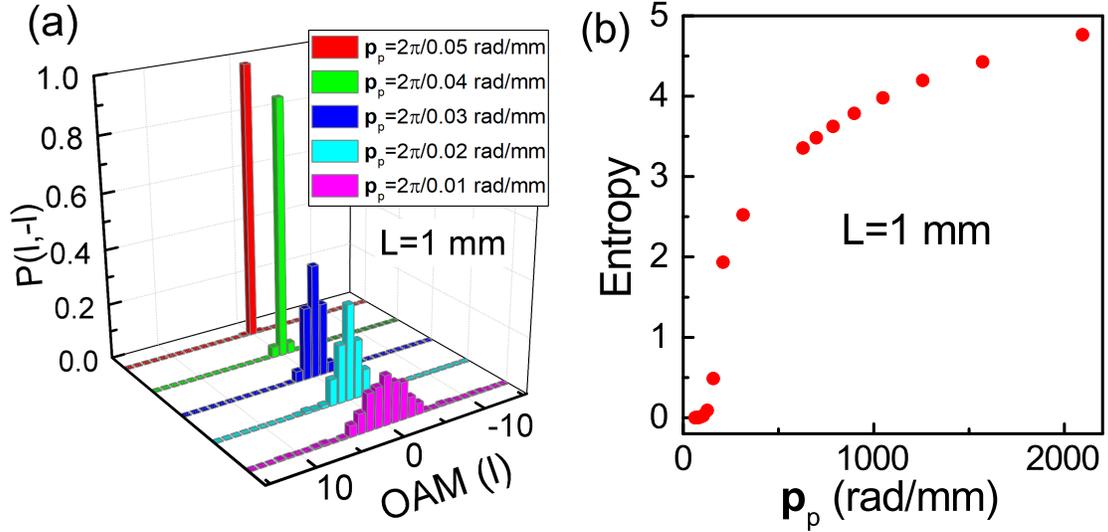}
    \caption{Dependence of the two-photon OAM entanglement on the radius of the pump cone sate, $\left| \mathbf{p}_{p} \right|$. (a) Probability distribution $P\left( l,-l \right)$ for the two-photon OAM entanglement for different pump cone states. (b) The dependence of the entanglement entropy on $\left| \mathbf{p}_{p} \right|$. In this simulation, the OAM of the pump state $l_{p}$ is set to $0$, and the length of the nonlinear crystal is fixed at $1mm$.}
    \label{fig:my_label}
\end{figure}

\par Fig.3 (a) shows the dependence of the mode probability distribution $P\left(l,-l\right)$ of the two-photon OAM entanglement on the radius of the pump cone state $\left|\mathbf{p}_{p}\right|$, in which $l$ ranges from -15 to 15 and $P\left(l,-l\right)$ is equal to
\begin{equation}
 P\left(l,-l\right)=\frac{\left|A\left(l,-l\right)\right|^{2}}{\sum_{l'}\left|A\left(l',-l'\right)\right|^{2}}. 
 \end{equation}
 Here we set $l_{p}=0$ and $L$ = 1 $mm$. Now the radius of the two-photon down-conversion cone $\Delta_{\mathbf{q}}$ is $2\pi/0.02$ $rad/mm$, and $A\left(l,-l\right)$ is a diagonal that crosses the zero point of $A\left( l_{s},l_{i} \right)$. From this figure, it can be seen that for the pump cone states with radii of $\left|\mathbf{p}_{p}\right|=2\pi/0.05 rad/mm$ and $2\pi/0.04 rad/mm$ which are smaller than the radius of the down-conversion two-photon cone $\Delta_{\mathbf{q}}$, the width of two-photon angular position correlation $\Delta\left(\theta_{s}+\theta_{i}\right)$ is much larger than 0, and $P\left(l,-l\right)$ is mainly concentrated near $l=0$. As $\left|\mathbf{p}_{p}\right|$ becomes comparable to or larger than $\Delta_{\mathbf{q}}$, the width of $P\left(l,-l\right)$ gradually rises, thus a higher dimensional OAM-entangled two-photon state arises.

\par The entanglement of a two-part system can be characterized by the so-called von Neumann entropy (or entanglement entropy) for the reduced state, which quantifies the number of entangled bits within the state. The entanglement entropy depends both on the entanglement dimension and degree. Fig.3 (b) shows the relationship between the von Neumann entropy for two-photon OAM entanglement and $\left|\mathbf{p}_{p}\right|$, where the entanglement entropy is expressed as \cite{PhysRevLett.84.5304, Miatto2012} 
\begin{equation}
E=-\sum_{l}P\left ( l,-l \right )\log_{2} P\left( l,-l \right ).
 \end{equation}
In accordance with the $\left|\mathbf{p}_{p}\right|$ dependence of OAM entanglement in Fig.3 (b), the entanglement entropy increases with $\left|\mathbf{p}_{p}\right|$. That is to say, as $\left|\mathbf{p}_{p}\right|$ increases, the entanglement dimension or degree or the both will increase. In the present simulation, because we limit the OAM to the range from -15 to 15, the entanglement entropy will finally saturate to 4.95. From this figure we can see that as $\left|\mathbf{p}_{p}\right|$ is larger than 2000 rad/mm, a near maximal two-photon entanglement with dimension of 31 can be obtained. It should be noted that before the steep increase of the entanglement entropy, there is a small non-rising region. This is due to the fact that for small values of $\left|\mathbf{p}_{p}\right|$ which is less than $\left|\mathbf{q}_{s(i)}\right|$, the approximation in equation (12) and (13) will not be fulfilled. In this region, the dependence of entanglement entropy on $\mathbf{p}_{p}$ is negligible and the down-converted two photons mainly occupy the $l=0$ mode.

\begin{figure}
    \centering
    \includegraphics[width=15cm]{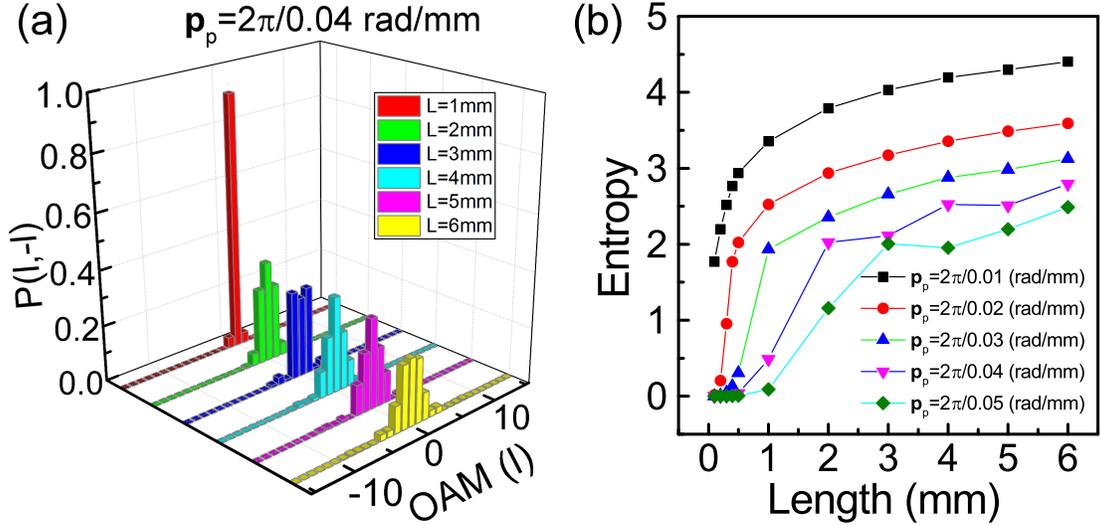}
    \caption{Dependence of the two-photon OAM entanglement on the length of the nonlinear crystal. (a) Probability distribution $P\left( l,-l \right)$ of the OAM entanglement for different crystal lengths. Here $\left| \mathbf{p}_{p} \right|=2\pi/0.004 rad/mm$ is taken for example. (b) Evolution of the entanglement entropy with respect to the length of the nonlinear crystal for different pump cone states.}
    \label{fig:my_label}
\end{figure}

\par Fig.4 (a) shows the dependence of the mode probability distribution $P\left(l,-l\right)$ of the two-photon OAM entanglement on the length of the nonlinear crystal $L$, where $\left|\mathbf{p}_{p}\right|$ is $2\pi/0.04 (rad/mm)$ and $l_{p}$ is $0$. Fig.4 (b) shows the dependence of the entanglement entropy on the crystal length for different pump cone states. From this figure we can see that the entanglement dimension and degree increase with $L$, the entanglement entropy gradually saturates at a rate depending on the $\left|\mathbf{p}_{p}\right|$. In contrast to the $\left|\mathbf{p}_{p}\right|$ dependence of the OAM entanglement which is ascribed to the increase of the pump cone radius, the $L$ dependence of OAM entanglement is due to the decrease in the radius of the two-photon down-conversion cone. From equation (5) it is predicted that the thicker the nonlinear crystal is, the less phase-mismatch the nonlinear parametric interaction can tolerate, thus the smaller the radius of the down-conversion cone is. In this case, the increase of $L$ will lead to a stronger angular position correlation $\Delta\left(\theta_{s}^{2}+\theta_{i}^{2}\right)$ (as shown in equation (13)). According to Heisenberg uncertainty principle, a wider OAM correlation bandwidth will be obtained\cite{pires2010measurement, jha2011partial}.

\par Similar to Fig.3 (b), there is also a non-rising region for the entanglement entropy in Fig.4. The width of this non-rising region increases with the decrease of $\left|\mathbf{p}_{p}\right|$. In this region, the radius of the two-photon down-conversion cone is larger than or comparable to the radius of the pump cone, then equation (13) will not be satisfied, the entanglement entropy is zero and the two-photon state mainly occupies the $l=0$ mode. Furthermore, both from Fig.(3,4) and equation (13), the increasing rate for the two-photon OAM entanglement with $L$ is slower than with $\left|\mathbf{p}_{p}\right|$.

\begin{figure}
    \centering
    \includegraphics[width=8cm]{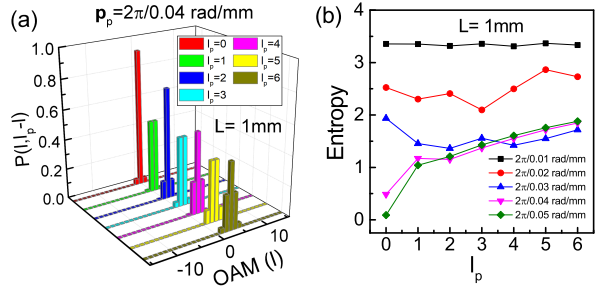}
    \caption{Dependence of the two-photon OAM entanglement on the OAM of the pump cone state. (a) Probability distribution $P\left( l,l_{p}-l \right)$ of the OAM entanglement for different $l_{p}$, the OAM carried by the pump state. The parameters for this simulation are $\left| \mathbf{p}_{p} \right|=0.05 rad/mm$ $L_{z}=1mm$. (b) Dependence of the entanglement entropy on $l_{p}$ for four different pump cone states.}
    \label{fig:my_label}
\end{figure}

\par Finally, we study the dependence of the OAM entanglement of on $l_{p}$, the OAM of the pump state. As shown in equation (10), the introduction of $l_{p}$ can be viewed as a displacement operation that shifts the two-photon OAM entanglement from $A\left(l,-l\right)$ to $A\left(l,l_{p}-l\right)$, the diagonals of $A\left(l_{s},l_{i}\right)$. The dependence of the mode probability distribution $P\left(l,l_{p}-l\right)$ of two-photon OAM entanglement is shown in Fig.5 (a), where $L$ and $\left|\mathbf{p}_{p}\right|$ are respectively of 1 $mm$ and 0.04 $rad/mm$. The mode probability distribution $P\left(l,l_{p}-l\right)$ centers at $l_{p}/2$, and the two-photon OAM entanglement varies with $l_{p}$. The corresponding dependence of the entanglement entropy on $l_{p}$ for different pump cone states is shown in Fig.5 (b). The entanglement entropy shows no explicit dependence on $l_{p}$. Compared with the $L$ dependence and $\left|\mathbf{p}_{p}\right|$ dependence of the OAM entanglement, the change of $l_{p}$ does not change the angular position correlation for the down-converted two photons, thus the change of $l_{p}$ is a less efficient way to enhance the OAM entanglement.

\par In the derivation of OAM entanglement in this paper, we have shown that two-photon angular position correlation and OAM correlation are tied closely. They form a EPR pair\cite{leach2010quantum} just like position and momentum\cite{howell2004realization}. The OAM entanglement can be viewed as unitary transformation operated on the two-photon angular position correlation, which is always used to quantify entanglement\cite{vedral1997quantifying}. The Heisenberg uncertainty principle can also be applied to correlated systems, that an increase of the correlation in one space will lead to a decrease in its conjugate space. But in present case, the pump state only carries a single OAM, thus the increase in angular position correlation will cause a increase in the OAM entanglement dimension and degree. For actual experiments in which an incident plane-wave like pump light with rotational symmetry is focused onto a nonlinear crystal, the angular position correlation can be strengthened by choosing a lens of shorter focal length and a thicker nonlinear crystal. In addition, to reduce the influence of the pump cone state near the central propagation axis, a ring-like pump beam may favor the generation of high-dimensional OAM entanglement.
\section{Conclusion}
\par In conclusion, we have given a variant theoretical interpretation of the down-converted two-photon OAM entanglement within the azimuthal region and pointed out several approaches to enhance entanglement dimension and degree. The pump state is decomposed into a set of pump cone states characterized by its radius. By fixing the transverse momentum projections of the signal and idler states, we discuss the two-photon correlation in azimuthal region. The entanglement dimension and degree, characterized by the entanglement entropy, show a strong dependence on the radius of the pump cone state and the length of the nonlinear crystal. Such phenomena can be explained by the Heisenberg uncertainty principle between angular position correlation and OAM correlation. In contrast, varying the OAM of the pump state just ends up with shifting the entanglement spectrum, which turns out to be a less efficient way to increase the OAM entanglement. 

\section{Acknowledgement}
This work is supported by Young fund of Jiangsu Natural Science Foundation of China (SJ216025), National fund incubation project (NY217024), Scientific Research Foundation of Nanjing University of Posts and Telecommunications (NY215034), National Natural Science Foundations of China (Grant No. 61475075).

\bibliography{reference}

\end{document}